\documentclass{aastex62}

\newcommand{\teff}{\mbox{T$_{\rm eff}$}}
\newcommand{\logg}{\mbox{log~{\it g}}}

\newcommand{\kmsec}{\mbox{km~s$^{\rm -1}$}}

\newcommand{\vsini}{\mbox{$v~{\rm sin}{\it i}$}}

\accepted{06/08/2018}

\shorttitle{Multiple populations in Galactic open clusters } 
\shortauthors{A.\,F. Marino, et al.} 

\begin{document}
\title{DISCOVERY OF EXTENDED MAIN SEQUENCE TURN OFFS IN GALACTIC OPEN CLUSTERS
}  

\author{A.\ F.\,Marino} 
\affiliation{Research School of Astronomy \& Astrophysics, Australian National University, Canberra, ACT 2611, Australia}  
\author{A.\ P.\,Milone}
\affiliation{Dipartimento di Fisica e Astronomia ``Galileo Galilei'' - Univ. di Padova, Vicolo dell'Osservatorio 3, Padova, IT-35122}
\author{L.\,Casagrande} 
\affiliation{Research School of Astronomy \& Astrophysics, Australian National University, Canberra, ACT 2611, Australia}  
\author{N.\ Przybilla}
\affiliation{Institut f\"{u}r Astro- und Teilchenphysik, Universit\"{a}t Innsbruck, Technikerstrasse 25, 6020, Innsbruck, Austria}
\author{L.\ Balaguer-N\'u\~nez}
\affiliation{Departament de Fisica Quantica i Astrofisica (Facultat de Fisica), Universitat de Barcelona, Marti i Franques 1, 08028, Barcelona, Spain}  
\affiliation{Institut d'Estudis Espacials de Catalunya (IEEC), C/Gran Capita, 2-4, E-08034, Barcelona, Spain}
\author{M.\ Di Criscienzo}
\affiliation{INAF - Osservatorio Astronomico di Roma, Monte Porzio, Italy}
\author{A.\ Serenelli}
\affiliation{Institut d'Estudis Espacials de Catalunya (IEEC), C/Gran Capita, 2-4, E-08034, Barcelona, Spain}
\affiliation{Institute of Space Sciences (ICE, CSIC) Campus UAB, Carrer de Can Magrans, s/n, E-08193, Barcelona, Spain}  
\author{F.\ Vilardell}
\affiliation{Institut d'Estudis Espacials de Catalunya (IEEC), C/Gran Capita, 2-4, E-08034, Barcelona, Spain}

\correspondingauthor{A.\ F.\,Marino}
\email{anna.marino@anu.edu.au}

\begin{abstract}  
By far, the color-magnitude diagrams (CMDs) of Galactic open clusters have
been considered proto-types of single stellar populations. 
By using photometry
in ultraviolet and optical bands we discovered that the nearby young
cluster NGC\,6705 (M\,11) exhibits an extended main-sequence turn off
(eMSTO) and a broadened main-sequence (MS). This is the first evidence of
multiple stellar populations in a Galactic open cluster. 
By using high-resolution VLT spectra we provide direct evidence that
the multiple sequences along the CMD correspond to stellar populations
with different rotation rates. 
Specifically, the blue MS is formed of
slow-rotating stars, while red-MS stars are fast rotators. 
Moreover, we exploit photometry from Gaia DR2 to show that three Galactic open
 clusters, namely NGC\,2099, NGC\,2360, and NGC\,2818, exhibit the
 eMSTO, thus suggesting that it is a common feature among these objects.

Our previous work on the Large Magellanic Cloud star cluster NGC\,1818
shows that slowly and rapidly-rotating stars 
populate the blue  and red MS observed in its CMD.
The similarities between M\,11 and the young clusters of the Magellanic Clouds suggest that
rotation is responsible for the appearance of
multiple populations in the CMDs of both Milky Way
open clusters and Magellanic Clouds young clusters.
\end{abstract}

\keywords{open clusters: individual (NGC\,6705) -- Hertzsprung-Russell diagram }

\section{Introduction}\label{sec:intro}

The Galactic open clusters are considered proto-types of
simple stellar populations. This assumption was supported by 
color-magnitude diagrams (CMDs) often well reproduced by single
isochrones (e.g.\,Kalirai et al.\,2003; Bedin et al.\,2010).  
In contrast, the CMDs of young and intermediate-age star clusters in
both Magellanic Clouds (MCs) are not 
consistent with single isochrones. 
Nearly all these clusters with ages between $\sim$700~Myrs and $\sim$2~Gyrs exhibit extended
main sequence turn off (eMSTO, e.g.\,Mackey \& Broby Nielsen\,2007;
Milone et al.\,2009; Goudfrooij et al.\,2011,\,2014); and 
clusters younger than $\sim$700~Myrs show both split main sequence (MS) and
eMSTO (e.g.\,Milone et al.\,2013,\,2015,\,2018; Correnti et
al.\,2015,\,2017; Bastian et al.\,2017; Li et al.\,2017).  

Stellar rotation plays a major role in the
split MS and the eMSTO. The comparison between the 
observed CMDs of young MCs clusters and isochrones
indicates that the blue MS (bMS) is consistent with a population of
non-rotating or slow-rotating stars, while red-MS (rMS) stars rotate
close to the critical value. According to this scenario, the
fainter part of the MSTO is mainly populated by fast rotators, while
the upper MSTO should host slow rotators (e.g.\,D'Antona et al.\,2015). 
This scenario is confirmed by the large fraction of Be stars in young
MCs clusters (e.g.\,Keller et al.\,2000; Bastian et
al.\,2017), mostly populating
the rMS and the faint MSTO (Milone et al.\,2018). 
Spectroscopic analysis of the Large Magellanic Cloud (LMC)
cluster NGC\,1866 has provided direct evidence that its eMSTO hosts
stars with different rotations, and that 
the faint MSTO is mostly populated by fast rotators (Dupree et
al.\,2017).
High-resolution spectra have shown that the split MS of the
LMC cluster NGC\,1818 hosts a bMS made of slow rotators, and a rMS
of stars with high rotation (Marino et al.\,2018). 

Stellar rotation is a common phenomenon that affects
many aspects of stellar evolution, in particular for stars with
$\mathcal{M}>$1.5-1.7~$\mathcal{M}_{\odot}$ (Meynet \& Maeder\,2000, and references
therein). Stars with rotation rates from slow to nearly
critical have been indeed reported both in field and open
cluster stars (e.g.\,Huang et al.\,2010), and the possible role of
stellar rotation in causing color spreads in some open clusters
was suggested by Brandt et al.\,(2015a,\,2015b). 

We combine Str\"omgren photometry and spectra of the
Galactic open cluster NGC\,6705 (M\,11) to investigate its MSTO and
upper MS. 
As estimated by Cantat-Gaudin et al.\,(2014), the age of this cluster
ranges from 250 to 320~Myr,
and its mass is between 3700 and 11000~$\mathcal{M}_{\odot}$.
We further exploit photometry, proper motions and
parallaxes from Gaia data release 2 (DR2, Gaia collaboration et
al.\,2018) to search
for multiple populations in Galactic open clusters NGC\,2099,
NGC\,2360, and NGC\,2818. 

\section{Data and analysis}\label{sec:data}

To search for multiple sequences along the CMD of M\,11 we used images
collected with the Isaac Newton Telescope (INT) through the $uvy$
filters as part of the Str\"omgren survey for asteroseismology and
Galactic archaeology (Casagrande et al.\,2014). 
Specifically, we used
10s$+$2$\times$20s$+$2$\times$50s$+$2$\times$120s $u$ band,  
 5s$+$2$\times$10s$+$2$\times$40s$+$2$\times$120s $v$ band, and
 3$\times$3s$+$2$\times$5s$+$2$\times$50s $y$ band images; 
 stellar photometry and proper motions are from Gaia DR2.  

The photometry and astrometry of the INT images was carried out using
{\sf kitchensync} (Anderson et al.\,2008) modified 
for INT images. 
For each exposure we derived a grid of 3$\times$3 empirical PSFs by using
the available isolated, bright, and non-saturated stars. We accounted for the
spatial variation of the PSF by assuming that each star of each image
is associated to a bi-linear weighted interpolation of the four
nearest PSFs (Anderson \& King\,2000). 

The fluxes of bright and faint stars have been derived by using
distinct procedures. Bright stars have been measured in each
exposure independently by using the best PSF model, and then 
averaged together. To derive fluxes and
positions of faint stars, 
poorly constrained in the single
exposures, we fitted all the pixels of all the exposures
simultaneously (see Anderson et al.\,2008).
The photometry has been calibrated as in Casagrande et al.\,(2014, see
their Sec.~2).  

Cluster members have been selected by using
proper motions and parallaxes from
Gaia DR2 (see Sec.~\ref{sec:phot_result}). 
The photometry of the cluster members has been corrected for differential
reddening as in Milone et al.\,(2012). 

The spectroscopic data consist of FLAMES/GIRAFFE data (Pasquini et
al.\,2002), with setup H665 ($R \sim \lambda/\Delta \lambda 
\sim$17000), collected under the Gaia-ESO Survey (Gilmore et al.\,2012) and
publicly available on the ESO archive\footnote{{\sf
    http://archive.eso.org/cms.html}}. 
From the available sample of stars observed in M\,11, we have selected
those in the magnitude range $12.0 \lesssim v \lesssim 14.5$, which is the
interval where the spread MS and MSTO is observed (see
Sec.~\ref{sec:phot_result}). 
At the H$\alpha$ line wavelength, the typical
signal-to-noise of the fully reduced spectra is $\sim$100.

Projected rotational velocities (\vsini) 
were determined by fitting the H$\alpha$ core. 
A grid of hybrid non-local thermodynamic
equilibrium (non-LTE) model spectra
was computed using the approach discussed by 
Przybilla \& Butler\,(2004a,b) based on {\sc Atlas9}
model atmospheres (Kurucz\,1993). Non-LTE level populations were
calculated using {\sc Detail} and synthetic spectra using {\sc
Surface} (Giddings\,1981; Butler \& Giddings\,1985, both updated by
K.\,Butler). 
The grid covers effective temperatures
\teff\ from 6400\,K to 8800\,K (step width of 200\,K),
with surface gravity fixed to \logg$=$4.0, which is the typical
value found in the Gaia-ESO survey for these stars. The consideration 
of non-LTE effects, yielding a pronounced strengthening of the 
H$\alpha$ Doppler core, is important for avoiding a systematic bias
of the \vsini\ 
in particular for slower rotators.
The model spectra were convolved with a Gaussian profile to account for 
instrumental broadening and a rotational profile 
\vsini\ from 0 to 300\,\kmsec\ (step width of 20\,\kmsec).

We employed a $\chi^{2}$ minimisation of each 
observed
spectrum around the H$\alpha$ line to simultaneously infer
\teff, primarily affecting the wings, and \vsini,
which has a major impact on the core.
The inferred \vsini\ values are listed in Table~1.

The choice of atmospheric parameters has little
impact on the \vsini\ determination. 
Varying \logg\ by $\pm$0.5~dex has negligible effect,
while changes in \teff\ by $\pm$200~K result in \vsini\ variations of
$\mp$20~\kmsec.
The major contribution to the errors 
is associated with 
the continuum placement, which introduces an internal
uncertainty of $\pm$30-40~\kmsec. 
Conservatively, we associate an internal error of $\pm$40~\kmsec\ to
our \vsini\ estimates.

%
   \begin{figure*}
   \centering
   \includegraphics[width=8.6cm,height=8.415cm]{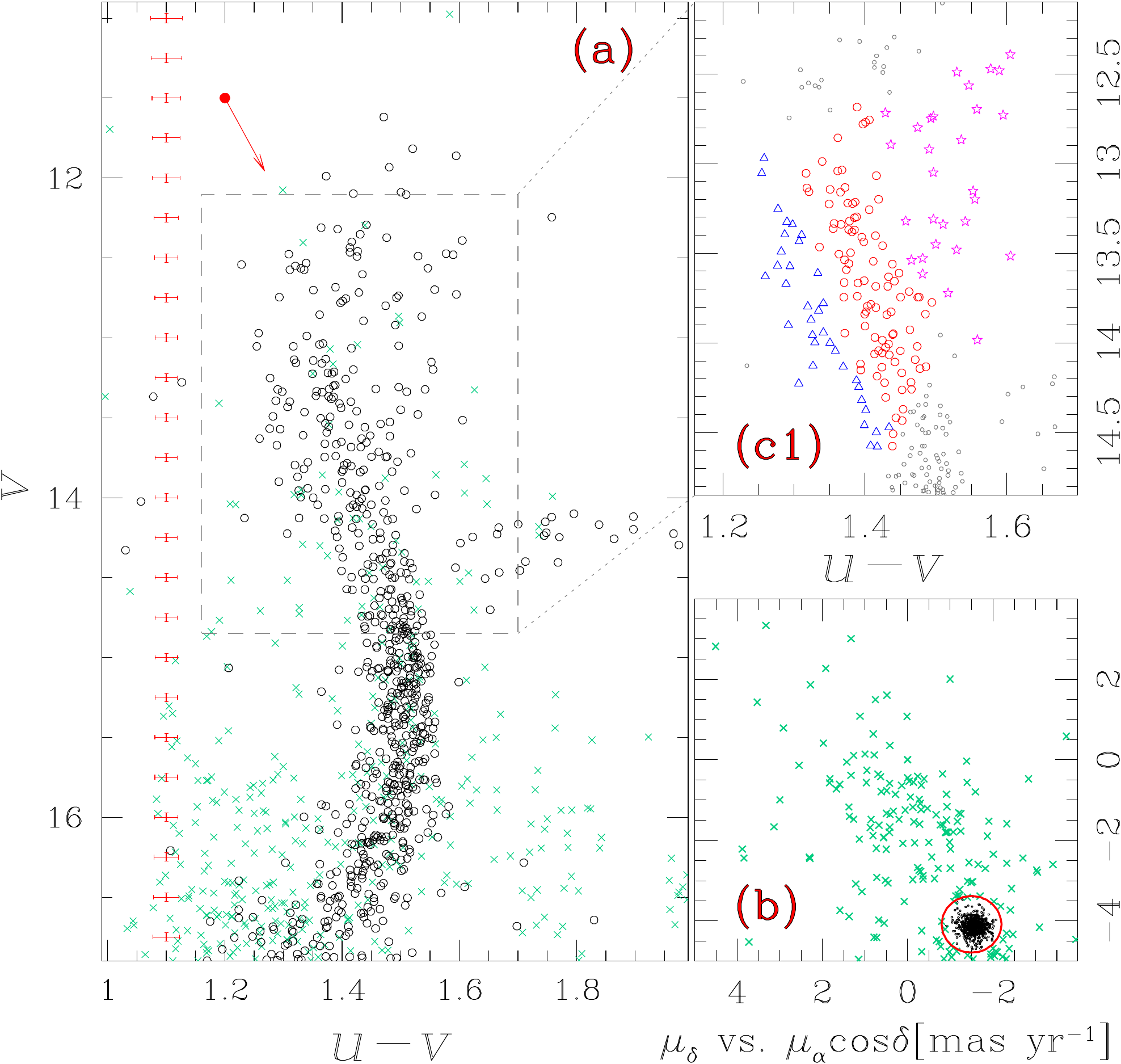}
   \includegraphics[width=8.6cm]{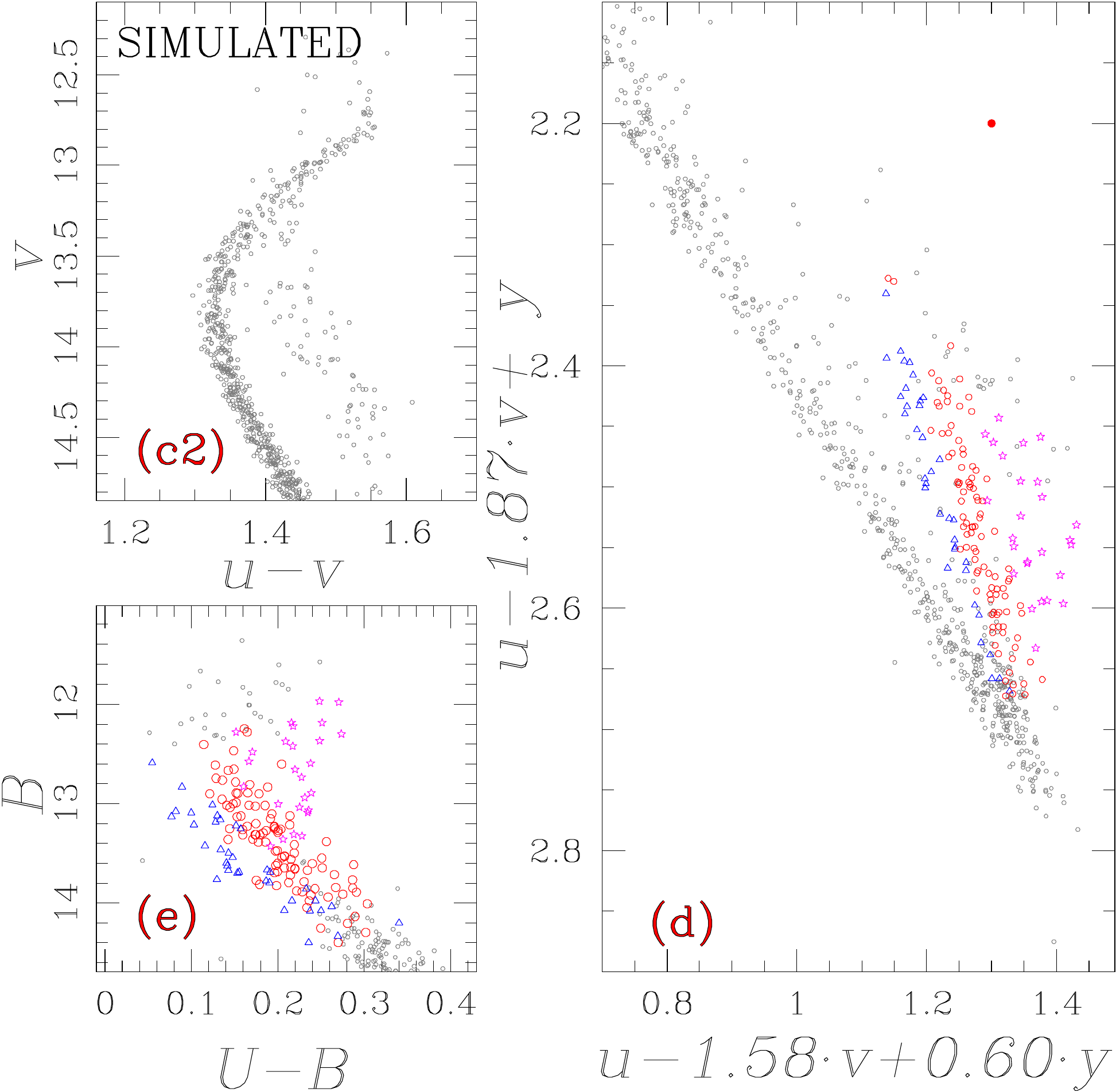}
   \caption{
     {\it Panel a}: $v$-$(u-v)$ CMD corrected for differential
     reddening of stars in the M\,11 field. Cluster members and field
     stars are plotted with black circles and aqua crosses,
     respectively. 
     The red arrow indicates the reddening
     vector. Typical photometric uncertainties are plotted on the
     left.
     {\it Panel b}: vector-point diagram of proper motions for the
     stars in the CMD. The red circle separates the bulk of cluster
     members from field stars. 
     {\it Panel c1}: Zoom-in of the CMD around the
     MSTO. bMS, rMS and eMSTO stars are
     represented with blue, red, and magenta symbols,
     respectively. The same colors are used in the other panels. 
     {\it Panel c2}: Synthetic $v$-$(u-v)$ CMD for one
       non-rotating population from a Geneva isochrone (Georgy
       et al.\,2013) with
       age$=$250~Myr, $Z=$0.014, $(m-M)_{\rm 0}=$10.85, $E(B-V)=$0.40.
     {\it Panel d}: Reddening-free diagram for cluster
     members.
     {\it Panels e}: Differential-reddening corrected $B$-$(U-B)$ CMD 
          (photometry by Sung et al.\,1999).} 
\label{fig:cmd}
 \end{figure*}
%

\section{The extended main-sequence turn off of M\,11}\label{sec:phot_result}

The $v$-($u-v$) CMD of stars in the M\,11 field is
plotted in Fig.~\ref{fig:cmd}a.
The vector-point diagram of proper motions plotted in
Fig.~\ref{fig:cmd}b reveals that M\,11 stars are clearly clustered
around ($\mu_{\alpha}cos{\delta}; \mu_{\rm \delta}$)=($\sim
-$1.57;$-$4.12)~mas~yr$^{-1}$, while the proper motions of field stars exhibit a broadened
distribution. 
As a further membership criterium, we estimated the median parallax of the proper-motion
selected stars ($<\omega>$=0.42~mas) and considered as members of M\,11 only
stars with parallaxes within 0.17~mas (corresponding to three
times the dispersion) from the median value.
We verified that the selected sample of cluster members matches the
M\,11 stars selected by Cantat-Gaudin et al.\,(2018).

The CMD of cluster members reveals that M\,11 hosts
an eMSTO and a broadened, possibly split, upper MS, in close analogy
with MC clusters at similar ages. 
The color and magnitude broadening of the eMSTO and of
the upper MS are significantly larger than those expected by
observational errors. 
The presence of these features is confirmed by the
synthetic $v$ vs.\,($u-v$) CMD (Fig.~\ref{fig:cmd}c2) corresponding
to one single non-rotating population, which does not show any
evidence of neither eMSTO nor spread MS.
The MS region with $v>14.5$ is consistent with a single stellar sequence. 

To further demonstrate that the eMSTO and the broadened MS are not
artifacts due to residual differential reddening we selected three
groups of bMS stars, rMS stars, and eMSTO stars by hand in the
$v$ vs.\,$u-v$ CMD plotted in Fig.~\ref{fig:cmd}c1. 
In Fig.~\ref{fig:cmd}d we plotted the magnitude
combination $u-1.87 \cdot v + y$ against $u -1.58 \cdot v + 0.60 \cdot
y$ for cluster members. This diagram is reddening-free 
by assuming for the M\,11 field of view 
the absorption
coefficients by Schlegel et al.\,(1998).
If the eMSTO and broadened MS are due to residual differential
reddening, we would expect that the three stellar groups exhibit random
values of the reddening-free quantities $u-1.87 \cdot v + y$ and $u
-1.58 \cdot v + 0.60 \cdot y$. 
The fact that the three groups of bMS, rMS,
and eMSTO stars are clearly separated in the reddening-free diagram 
demonstrates that eMSTO and broadened MS are intrinsic features of M\,11. 
Panels (e) shows the $B$ vs.\,($U-B$) CMD from Sung et al.\,(1999)
corrected for differential reddening. 
The selected bMS, rMS, and eMSTO stars populate distinct regions of
this diagram thus corroborating the presence of multiple stellar
populations in M\,11.

\section{Rotation}\label{sec:rotation}

\begin{figure*}
   \centering
   \includegraphics[width=7.62cm]{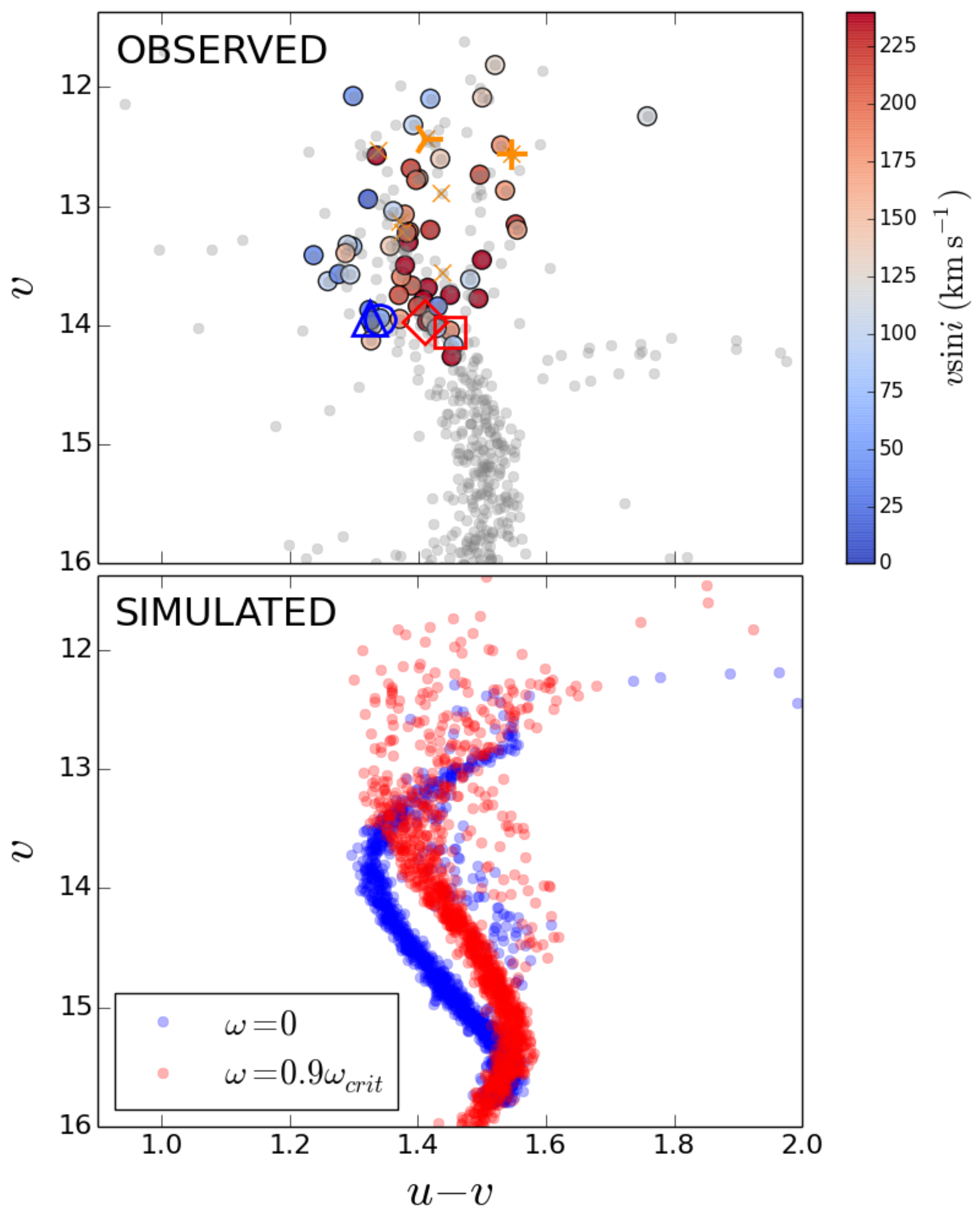}
   \includegraphics[width=9.5cm]{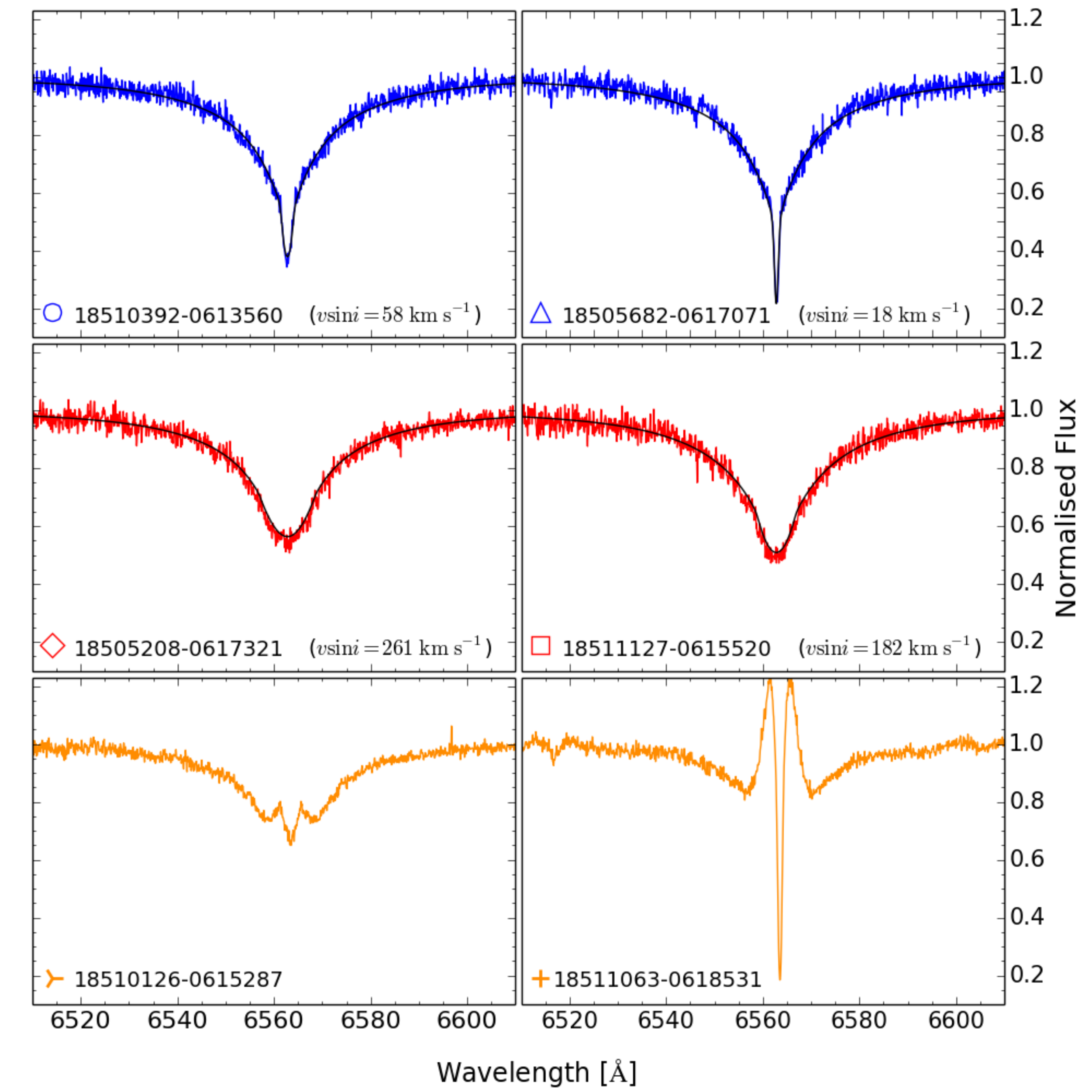}
      \caption{{\it Left-upper panel}: 
        observed $v$-$(u-v)$ CMD of M\,11. 
        Spectroscopically analyzed stars are marked with large circles. The
        color of each circle is associated with the corresponding
        value of \vsini\ as indicated by the color-scale on the
        right. 
        {\it Left-lower panel}: $v$-$(u-v)$ synthetic CMD
          corresponding to two coeval populations (age$=$250~Myr),
          one with no rotation (blue) and the other with high rotation
          ($\omega$=0.9$\omega_{crit}$, red). The expected photometric
        error has been added to the simulations
        {\it Right panels}: some examples of our spectra.
        From the upper to lower panels we show two bMS and two rMS at
        similar magnitude, and two H$\alpha$ core emission stars.
        For each bMS and rMS spectrum, 
        we super-impose to the observed spectra the
        best-fitting non-LTE model in black.
       }
        \label{fig:MSTO_CMD}
   \end{figure*}

The position of our spectroscopically analyzed stars on the
$v$-$(u-v)$ CMD is shown in the left-upper panel of Fig.~\ref{fig:MSTO_CMD}. 
They are distributed along the eMSTO and the broadened MS of M\,11, discussed in
Sec.~\ref{sec:phot_result}. The side-bar 
shows the color scale indicative of the inferred \vsini.
Stars exhibiting a H$\alpha$ emission core have been clearly
indicated. Our projected rotational velocities results suggest that 
\vsini\ values span a large range, from a
few tens to more than 250~\kmsec, 
and that stars with slower rotation are distributed on the bluer side
of the CMD.  

For comparison reasons, the synthetic CMD of
Fig.~\ref{fig:cmd}c2 for a no-rotating stellar
population has been reproduced in the left-bottom panel of
Fig.~\ref{fig:MSTO_CMD}. To this no-rotating population we
have added a synthetic coeval rapidly-rotating
population, with $\omega$=0.9$\omega_{crit}$
(Georgy et al.\,2013). If a
rotation regime close to the critical velocity can account for a
spread in the MSTO, the split MS can only be reproduced by two
stellar populations with intrinsic different stellar rotation rates. 

The right panels of Fig.~\ref{fig:MSTO_CMD} illustrate some examples
of our spectral fitting. 
Specifically, upper and middle panels show
two bMS and rMS stars, respectively, chosen to have similar luminosities.
The best-fit synthetic spectra are super-imposed to each spectrum.
The lower panels display the spectra of two H$\alpha$-emission 
stars in our sample, for which we could not infer any \vsini. 
A visual inspection of these spectra immediately suggests a different
\vsini\ for the bMS and rMS stars, having similar luminosity, but
different color.
The spectral profiles of the two rMS stars are significantly broader
than those of the bMS ones.

\begin{figure*}
   \centering
     \includegraphics[width=18cm]{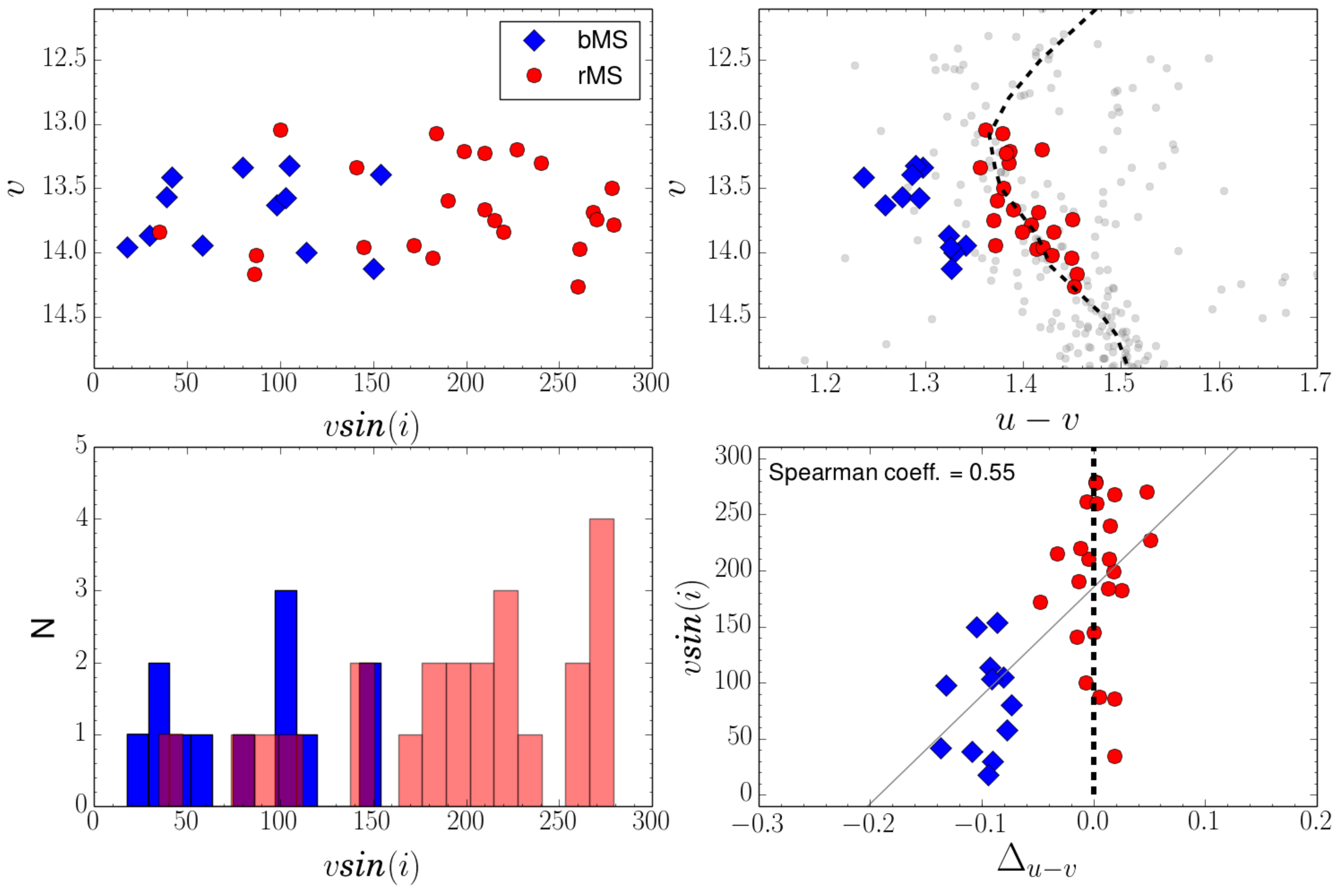}
      \caption{
{\it Left}: $v$ mag vs.\,\vsini\ for 
MS stars in M\,11 (upper panel); 
on the lower panel we show the histogram distribution of \vsini\ for
rMS (red) and bMS (blue).
{\it Right}: $v$-$(u-v)$
CMD zoomed on the MS. The black-dashed line is a
fiducial for the rMS (upper panel). 
The lower panel shows \vsini\ as a function of the difference between
the $(u-v)$ color of each MS star and the
color of the fiducial ($\Delta_{u-v}$). The
black-dashed line corresponds to $\Delta_{u-v}$=0.
In all the panels rMS and bMS are represented with red-filled
circles, and blue diamonds, respectively. 
       }
        \label{fig:rMS_bMS}
   \end{figure*}

Figure~\ref{fig:rMS_bMS} is a closer look at MS.
The location on the CMD of the MS stars with available spectroscopy 
is displayed for stars with $v~\gtrsim$13.
Our targets are clearly distributed on the split MS of M\,11,
and the association with the bMS and rMS is straightforward (right-upper panel).

The \vsini\ histogram distributions, shown
in the lower-left panel, illustrate the quite wide range for both the bMS and rMS.
Blue-MS stars do not show any value $\gtrsim$150~\kmsec.
As \vsini\ values are lower limits to the real stellar rotation,
depending on the inclination, more likely, a fast-rotating star has a
higher \vsini\ than a slow rotator.
Although a certain degree of overlap in \vsini\ between bMS and rMS 
might be expected because of projection effects, we cannot exclude
some slower rotators among the rMS.  
 
The left-upper panel of Fig.~\ref{fig:rMS_bMS} displays the $v$ magnitude versus \vsini\
for rMS and bMS stars.
Clearly, bMS stars are slower rotators, while the rMS hosts stars with
higher rotation. In the right panels we analyze the color of
the MS stars as a function of \vsini.
To this aim we have drawn a fiducial line by eye defining the rMS on
the $v$-$(u-v)$ CMD (upper-right panel). 
The difference in color between each MS star and the fiducial, $\Delta_{u-v}$, does not show any
significant trend with \vsini\ (lower-right panel).  

The average \vsini\ we obtain for the 12 bMS and the 23 rMS are 
$<$\vsini$>_{\rm bMS}$=83$\pm$14~\kmsec\ ($\sigma$=46~\kmsec) and
$<$\vsini$>_{\rm rMS}$=194$\pm$15~\kmsec\ ($\sigma$=68~\kmsec),
respectively, meaning that a
difference in the rotation regimes of the two MSs exists.
The mean \vsini\ difference is $<\Delta_{\rm {\vsini_{rMS}-\vsini_{bMS}}}>$=111$\pm$21~\kmsec.  

%

%
%

\section{Extended main-sequence turn off as a common feature of open clusters. The cases of NGC\,2099, NGC\,2360, and NGC\,2818.}

\begin{figure*}
   \centering
   \includegraphics[width=8cm]{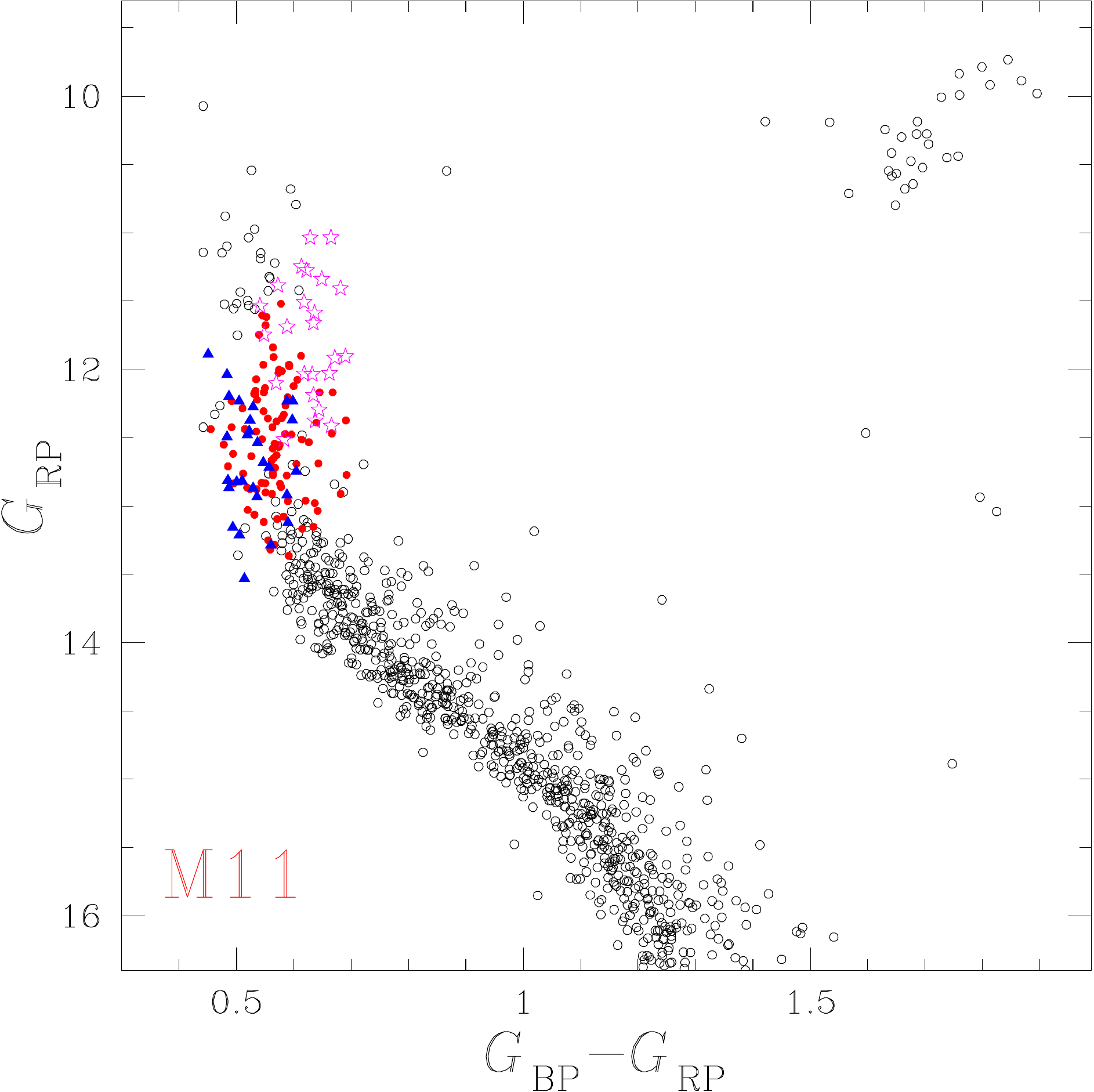}
   \includegraphics[width=8cm]{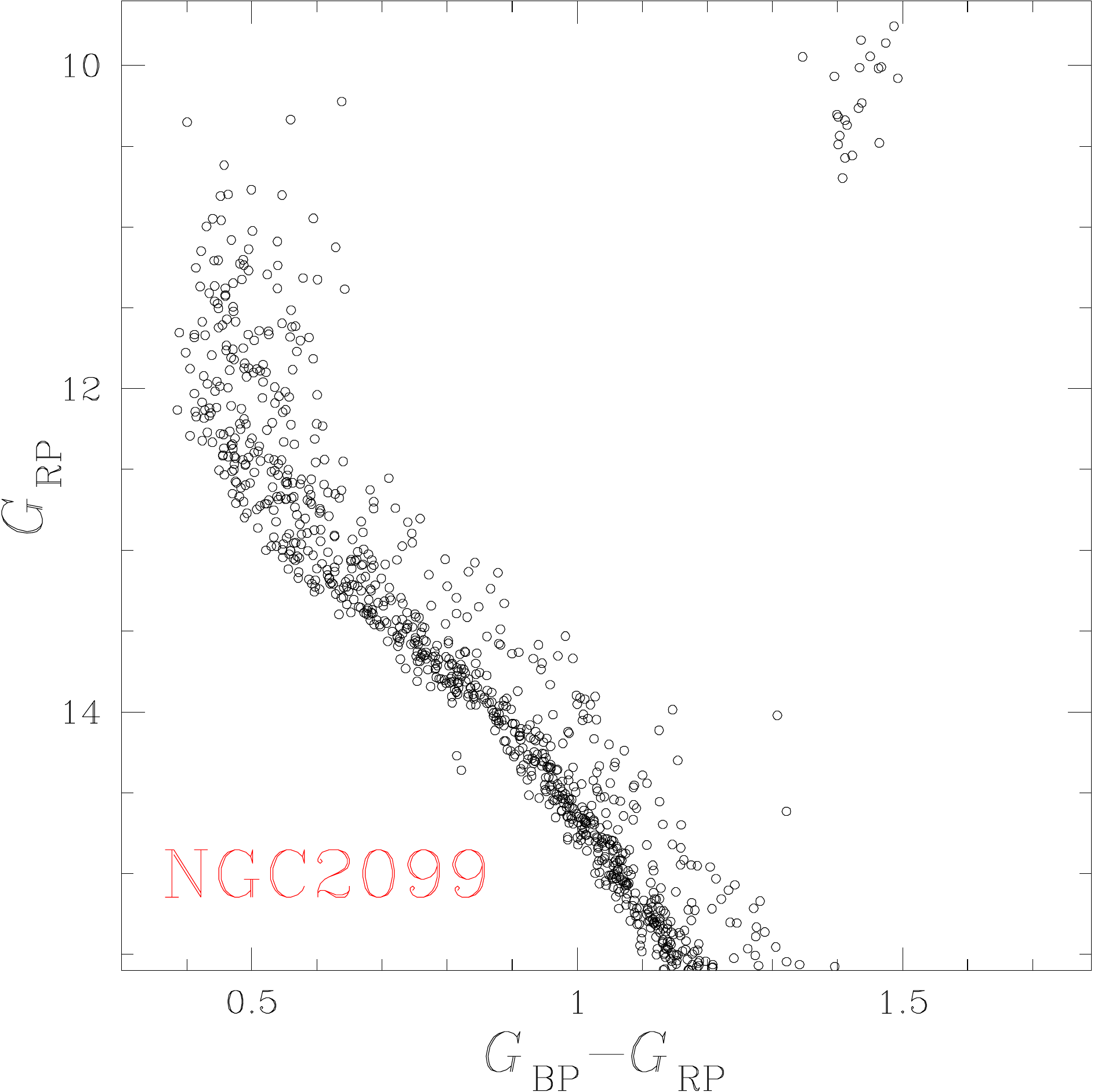}
   \includegraphics[width=8cm]{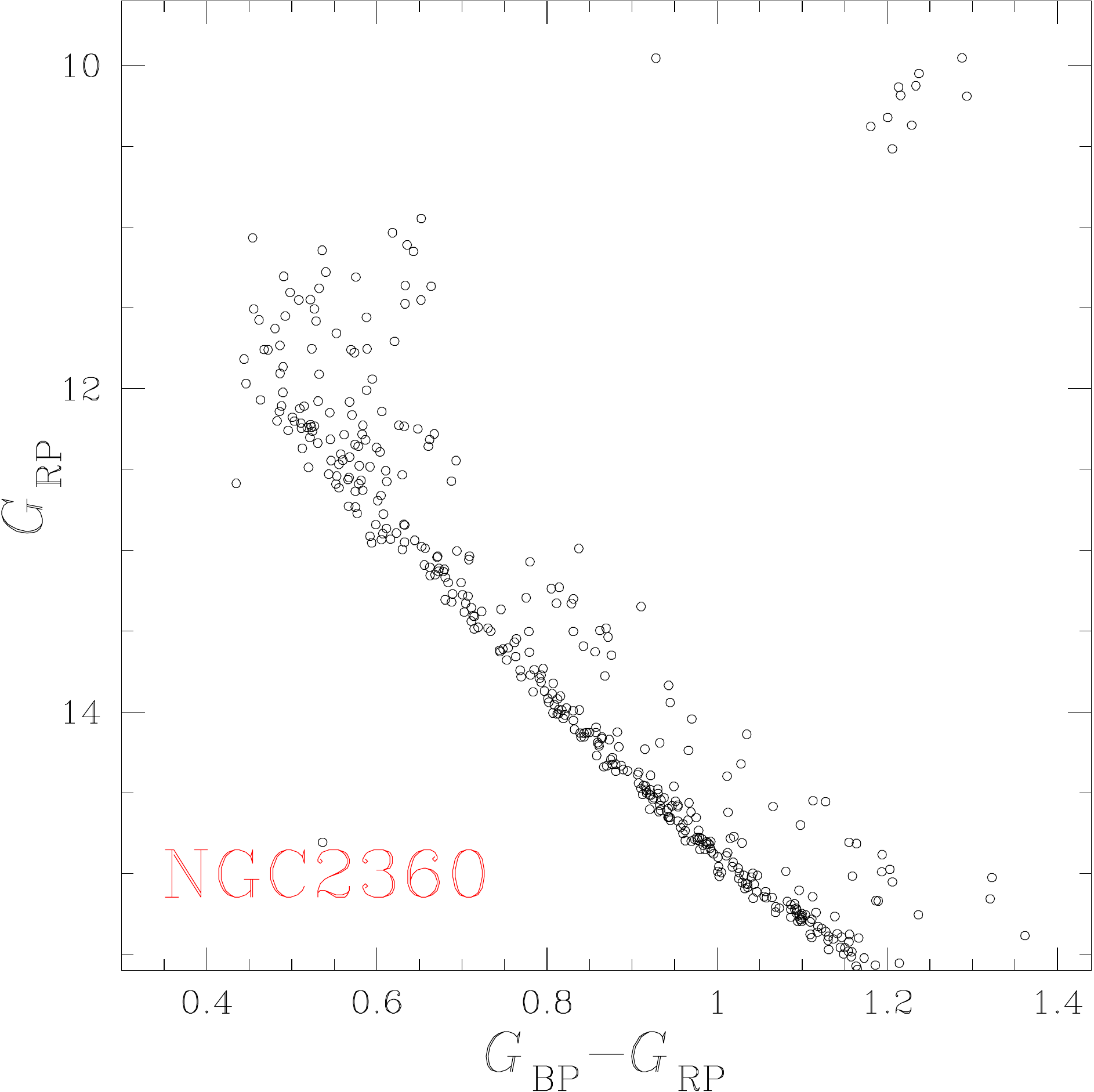}
   \includegraphics[width=8cm]{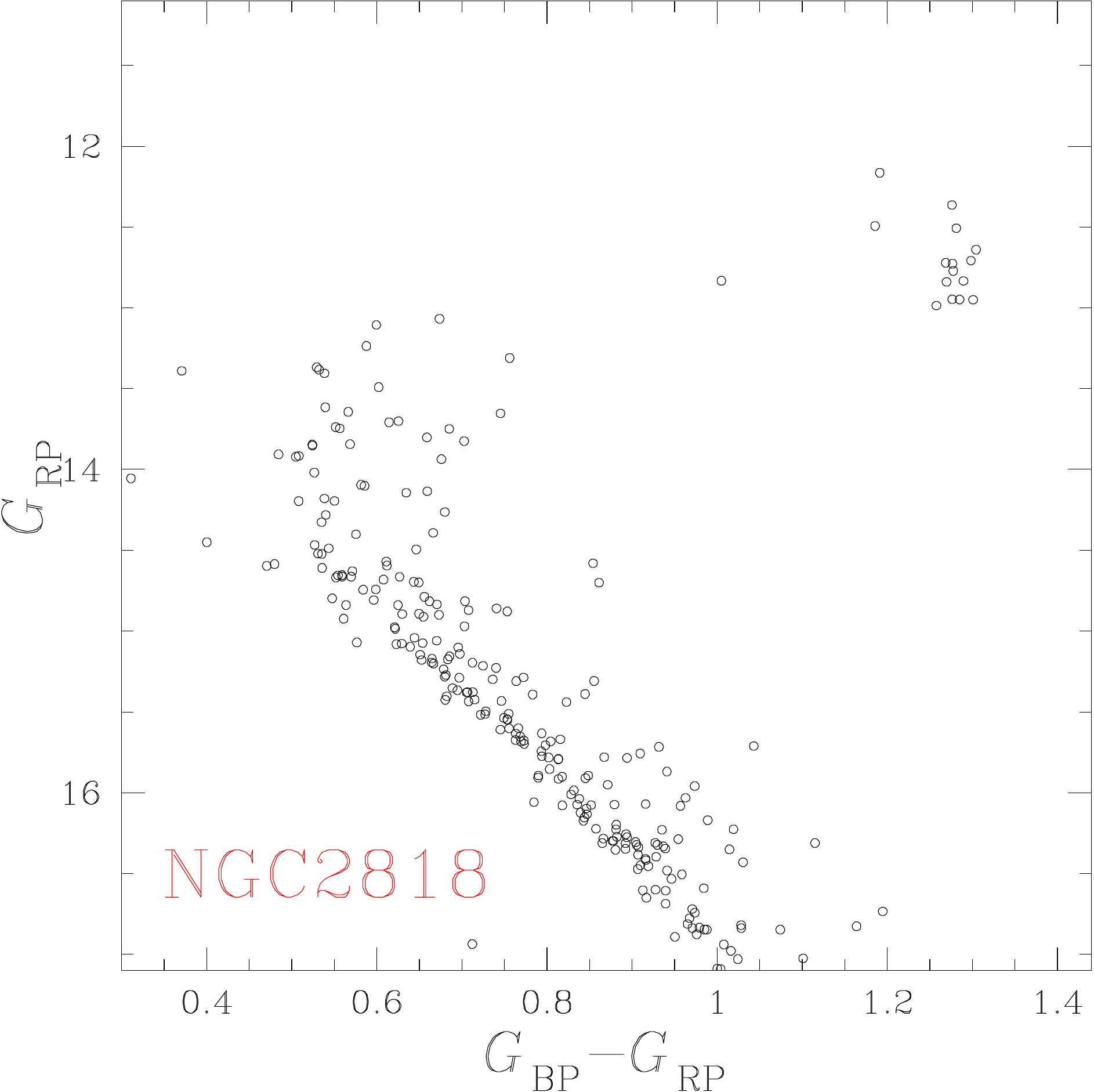}
      \caption{CMDs from Gaia DR2 photometry of Galactic open
        M\,11, NGC\,2099, NGC\,2360, and NGC\,2818,
        exhibiting a eMSTO. The three groups of bMS, rMS, and
        eMSTO stars of M\,11, selected in Fig.~1, are colored blue,
        red, and magenta, respectively. The binaries sequence is clearly distinguishable
      in the CMDs of NGC\,2099, NGC\,2360 and NGC\,2818.}
        \label{fig:Gaia}
   \end{figure*}

Ultraviolet and optical filters are efficient tools to identify
multiple sequences in the CMDs of young clusters. 
Although in optical filters the split MS and the eMSTO are less
prominent than in the ultraviolet, these features can be detected also in 
purely optical photometric diagrams (e.g.\,Milone et al.\,2013,\,2016;
Correnti et al.\,2015).  
As an example, in the upper-left panel of Fig.~\ref{fig:Gaia}, we plot
the $G_{\rm RP}$ vs.\,$G_{\rm BP}-G_{\rm RP}$ CMD of M\,11 cluster
members from Gaia DR2 photometry. 
The three groups of bMS, rMS and eMSTO stars exhibit different colors,
further confirming that the CMD of M\,11 is not consistent with a
single isochrone, and demonstrating that Gaia photometry is able to
detect eMSTOs. 

Driven by this result, we started inspecting the Gaia DR2 database to
search for multiple populations in open clusters. 
Our first results are shown in Fig.~\ref{fig:Gaia}, where we plot the 
CMDs for three Galactic open clusters, namely NGC\,2099, NGC\,2360, and NGC\,2818. 
As for M\,11, cluster members were selected from Gaia DR2 proper motions and parallaxes, and 
the CMDs were corrected for differential reddening. Noticeably, these clusters exhibit an eMSTO.
The MS is much narrower than the eMSTO and its color spread is
generally consistent with observational errors, thus demonstrating
that the eMSTO is not due to photometric uncertainties or residual
differential reddening.  

The analysis of a large sample of MCs clusters with ages of
$\sim$40\,Myr-2.5\,Gyr, has demonstrated that eMSTOs are 
common among these objects, while split MSs are observed in clusters
younger than $\sim$700~Myr (e.g.\,Milone et al.\,2009,\,2018), among
stars with masses $\gtrsim 1.6 \mathcal{M}_{\odot}$. Such mass
corresponds to the MS kink associated to the onset of the envelope
convection due to the opacity peak of partial H ionization
(e.g.\,D'Antona et al.\,2003).

Although a eMSTO is displayed by all the clusters shown in
Fig.~\ref{fig:Gaia}, remarkably a broaden MS is present only in
M\,11 and NGC\,2099.
By comparing the CMDs of Fig.~\ref{fig:Gaia} with isochrones from Marigo et
al.\,(2017), we find that these two clusters
have ages of $\sim$300~Myr
$\sim$500~Myr, respectively; NGC\,2818 and NGC\,2360
are well fitted by
$\sim$800~Myr and $\sim$1.1~Gyr isochrones.  
This fact corroborates the conclusion that the eMSTOs and
broaden/split MSs 
observed in the Milky Way clusters 
originate from the same physical mechanism responsible for the
appearance of similar features in the MCs clusters. 

%
\section{Conclusions}\label{sec:discussion}

 The eMSTO has been first discovered as a common feature in the CMD of
 the Large and Small MCs clusters younger than
 $\sim$2.5~Gyr (e.g.\,Mackey \& Broby Nielsen\,2007; Glatt et
 al.\,2008; Milone et al.\,2009,\,2018; Goudfrooij et al.\,2011,\,2014).  
This finding and the evidence of split MSs 
in MC clusters younger than $\sim 800$~Myrs (e.g.\,Milone et al.\,2015;
2018; Correnti et al.\,2017; Bastian et al.\,2017) has challenged the
theories on the formation and evolution of these objects. 

The eMSTO has been first interpreted as the signature of an extended
star-formation history, with young MC clusters considered as the
counterparts of the old GCs with multiple 
populations (e.g.\,Conroy et al.\,2011; Keller et al.\,2011). 
Alternatively, the eMSTO is due to coeval
stellar populations with different rotation (e.g.\,Bastian \& De\,Mink\,2009). 

Recent studies have provided strong evidence that rotation plays a major role in shaping
the split MS and the eMSTO. 
Direct evidence that the rMS is populated by fast rotators while the
bMS hosts stars with low rotation comes from spectroscopic
measurements of MS stars in the LMC cluster
NGC\,1818 (Marino et al.\,2018), while evidence on the connection
between the eMSTO and stars with different rotation rates has been provided
for NGC\,1866 (Dupree et al.\,2017).   

We have combined $uvy$ Str\"omgren photometry and Gaia DR2 proper motions and
parallaxes to search for multiple
populations in the Galactic open cluster M\,11. The $v$ vs.\,$(u-v)$
CMD shows an eMSTO and a broadened,
possibly split, MS in the magnitude interval $12.0 \lesssim v \lesssim
14.5$. 
We have demonstrated that these are intrinsic
features of the M\,11 CMD, in close analogy with what was previously observed
in nearly all the young clusters of both MCs. 
This is the first evidence of 
a connection between the eMSTO in MCs and in Galactic open clusters. 

We have exploited GIRAFFE data to derive projected rotational velocities 
for a sample of stars in M\,11.
Our analysis shows a clear difference in the mean \vsini\
between bMS and rMS stars, having average values of $<$\vsini$>_{\rm
  bMS}$=83$\pm$14~\kmsec\ ($\sigma$=46~\kmsec) and 
    $<$\vsini$>_{\rm rMS}$=194$\pm$15~\kmsec\ ($\sigma$=68~\kmsec). 
A large range in \vsini\ is also present among 
eMSTO stars, with some of them showing H$\alpha$ emission.

We show that the eMSTO of M\,11 is visible in the $G_{\rm RP}$
vs.\,$G_{\rm BP}-G_{\rm RP}$ CMD from Gaia DR2 
photometry. Driven by this finding, we started to search for multiple
populations in Galactic open clusters by using this dataset 
and found eMSTOs in NGC\,2099, NGC\,2360, and NGC\,2818. 

Our results indicate that the eMSTO is not a peculiarity of the
extragalactic 
MC clusters but is a common feature of Galactic open
clusters, thus challenging the traditional idea that these objects are
the proxy of single isochrone. 
High-resolution spectroscopy has provided direct evidence that stellar
rotation is the major responsible for the split MS and the eMSTO in
open clusters, making these objects similar to MCs clusters. 

\begin{deluxetable}{ccccc}
\tabletypesize{\scriptsize}  
\tablewidth{0pt}
\tablecaption{Identifiers from the Gaia-ESO survey, projected
  rotational velocities (\vsini) and photometry ($v$-$(u-v)$) for our spectroscopic targets.\label{tab:data}}
\tablehead{
  ID           & \vsini\ & $v$ & $u-v$  & CMD~region    \\
  (Gaia-ESO)   & [\kmsec] &     & mag&\\
}
\startdata
    18510414-0616202  &     270   &   12.57   &    1.34 & eMSTO \\ 
    18510358-0616573  &     178   &   12.77   &    1.40 & eMSTO \\ 
    18511514-0614431  &     212   &   12.74   &    1.50 & eMSTO \\ 
    18505981-0615291  &     138   &   12.09   &    1.50 & eMSTO \\ 
    18505829-0615284  &     139   &   12.60   &    1.43 & eMSTO \\ 
    18510593-0614348  &     130   &   11.82   &    1.52 & eMSTO \\ 
    18505407-0616503  &      76   &   12.10   &    1.42 & eMSTO \\ 
    18510656-0614562  &      47   &   12.08   &    1.30 & eMSTO \\ 
    18510793-0617217  &      99   &   12.32   &    1.39 & eMSTO \\ 
    18505098-0615314  &     176   &   12.87   &    1.54 & eMSTO \\ 
    18505254-0617374  &     180   &   12.49   &    1.53 & eMSTO \\ 
    18510577-0615230  &     220   &   12.68   &    1.39 & eMSTO \\ 
    18505436-0614545  &     202   &   12.78   &    1.40 & eMSTO \\ 
    18510462-0616124  &      17   &   12.94   &    1.32 & eMSTO \\ 
    18511517-0615541  &     117   &   12.25   &    1.76 & eMSTO \\ 
    18510512-0614075  &      30   &   13.87   &    1.32 & bMS \\   
    18510368-0617353  &      42   &   13.41   &    1.24 & bMS \\   
    18510392-0613560  &      58   &   13.94   &    1.34 & bMS \\   
    18510407-0618579  &     114   &   14.00   &    1.33 & bMS \\   
    18505693-0616214  &      80   &   13.34   &    1.30 & bMS \\   
    18505473-0615364  &      98   &   13.63   &    1.26 & bMS \\   
    18510907-0618579  &     105   &   13.32   &    1.29 & bMS \\   
    18510214-0616502  &     154   &   13.39   &    1.29 & bMS \\   
    18511287-0617194  &     150   &   14.13   &    1.33 & bMS \\   
    18505682-0617071  &      18   &   13.96   &    1.33 & bMS \\   
    18511086-0616295  &      39   &   13.57   &    1.28 & bMS \\   
    18511441-0614423  &     103   &   13.57   &    1.29 & bMS \\   
    18510255-0614488  &     240   &   13.30   &    1.39 & rMS \\   
    18505208-0617321  &     261   &   13.97   &    1.41 & rMS \\   
    18511127-0615520  &     182   &   14.04   &    1.45 & rMS \\   
    18510452-0615406  &     268   &   13.68   &    1.42 & rMS \\   
    18510092-0618029  &     172   &   13.95   &    1.37 & rMS \\   
    18510099-0616523  &     184   &   13.07   &    1.38 & rMS \\   
    18505944-0618212  &     210   &   13.67   &    1.39 & rMS \\   
    18510522-0615219  &     145   &   13.96   &    1.42 & rMS \\   
    18510572-0617177  &     199   &   13.21   &    1.39 & rMS \\   
    18510687-0617537  &     190   &   13.59   &    1.37 & rMS \\   
    18505345-0616096  &     215   &   13.75   &    1.37 & rMS \\   
    18510143-0617510  &     270   &   13.74   &    1.45 & rMS \\   
    18510176-0614073  &     100   &   13.04   &    1.36 & rMS \\   
    18510737-0618226  &     279   &   13.79   &    1.41 & rMS \\   
    18511195-0618463  &     278   &   13.50   &    1.38 & rMS \\   
    18505296-0617402  &      87   &   14.02   &    1.43 & rMS \\   
    18510286-0615250  &      86   &   14.17   &    1.46 & rMS \\   
    18505244-0618002  &     227   &   13.20   &    1.42 & rMS \\   
    18510891-0616433  &     210   &   13.22   &    1.38 & rMS \\   
    18505573-0617293  &     220   &   13.84   &    1.40 & rMS \\   
    18511094-0615043  &      35   &   13.84   &    1.43 & rMS \\   
    18511164-0618114  &     260   &   14.26   &    1.45 & rMS \\   
    18511196-0619220  &     141   &   13.33   &    1.36 & rMS \\   
    18510126-0615287  &      --   &   12.44   &    1.41 & H$\alpha$~emission \\  
    18510488-0614370  &      --   &   12.89   &    1.44 & H$\alpha$~emission \\  
    18505844-0613451  &      --   &   12.53   &    1.34 & H$\alpha$~emission \\  
    18511063-0618531  &      --   &   12.56   &    1.55 & H$\alpha$~emission \\  
    18505950-0615397  &      --   &   13.56   &    1.44 & H$\alpha$~emission \\  
    18505883-0616295  &      --   &   13.22   &    1.38 & H$\alpha$~emission \\  
    18505797-0615472  &      --   &   13.13   &    1.37 & H$\alpha$~emission \\  
\enddata
\end{deluxetable}

\acknowledgments
The authors thank the referee for her/his insightful comments.
AFM and LC acknowledge support by the Australian Research Council through
Discovery Early Career Researcher Award DE160100851 and the Future Fellowship FT160100402. 
APM has been supported by the European Research Council through the
Starting Grant ``GALFOR'' (716082) and the FARE-MIUR project R164RM93XW ``SEMPLICE''.
AS, LBN, and FV are partially supported by the MINECO (Spanish Ministry of
Economy) through grants ESP2017-82674-R and ESP2016-80079-C2-1-R (MINECO/FEDER, UE), SGR-1131 
(Generalitat Catalunya), and MDM-2014-0369 of ICCUB (Unidad de Excelencia 'Mar\'i\a de Maeztu')

\bibliographystyle{aa}

\end{document}